\documentclass[aps,pra,12pt]{revtex4}
\usepackage{graphicx}
\usepackage{amsmath}
\usepackage{amsfonts}

\usepackage{times}
\begin{document}

\title{Non-abelian anyons from degenerate Landau levels of ultracold atoms 
in artificial gauge potentials}

\author{Michele Burrello}
\affiliation{SISSA and INFN, Sezione di Trieste, 
Via Bonomea 265, I-34136, Trieste, Italy}

\author{Andrea Trombettoni}
\affiliation{SISSA and INFN, Sezione di Trieste, 
Via Bonomea 265, I-34136, Trieste, Italy}

\date{\today}

\begin{abstract}
We show that non-abelian potentials acting on ultracold gases 
with two hyperfine levels can give rise to ground states 
with non-abelian excitations. We consider a 
realistic gauge potential for which the Landau levels can be 
exactly determined: the non-abelian part of the vector potential 
makes the Landau levels non-degenerate. In the presence of strong 
repulsive interactions, deformed Laughlin ground states occur in general. 
However, at the degeneracy points of the Landau levels, 
non-abelian quantum Hall states appear:
these ground states, including deformed 
Moore-Read states (characterized by Ising anyons as quasi-holes), 
are studied for both fermionic and bosonic gases. 
\end{abstract}

\maketitle

\vspace{1.5cm}

The quest for states of matter having non-abelian excitations 
has been very intense in the past two decades, motivated by 
interest for the topological properties of such correlated 
states \cite{stern08,nayak08} and by their relevance for the 
implementation of topological quantum computation schemes \cite{nayak08}. 
Quantum Hall systems provide a major arena in which investigating 
non-abelian anyons, and a huge amount of work has been devoted 
to characterize which quantum Hall states are non-abelian and 
to propose (mainly interferometric) experiments to test the 
non-abelian nature of such states 
\cite{nayak08}. Although indirect evidences has been found 
for the $\nu=5/2$ quantum Hall state \cite{dolev08,west}, a completely unambiguous 
signature of its non-abelian nature is still lacking and the experimental manipulation 
of anyons in such systems stands so far as an important open challenge \cite{nayak08}. 

A significant and complementary issue consists in finding other 
experimental systems suitable to simulate the quantum Hall physics and realize 
non-abelian anyons. In this respect, ultracold atomic systems provide 
a natural candidate \cite{cooper01,aguado08}, 
due to the possibility of using them 
as simulators of many-body systems \cite{bloch08}. Two essential (but 
in general not sufficient) ingredients 
are available in ultracold atomic systems. 
First, one can simulate artificial magnetic fields by using rotating 
traps \cite{cooper08} or spatially dependent optical couplings 
between internal states of the atoms \cite{lin09}. 
Correspondingly, 
the possibility to have Laughlin states (with abelian excitations) 
using strongly interacting, 
rapidly rotating two-dimensional ultracold Bose gases has been discussed 
in the literature and it has been showed as well that 
incompressible vortex liquid states obtained for rotating ultracold bosons 
can be described by the Moore-Read wavefunction \cite{cooper01} 
(see the review \cite{cooper08}). 
The second ingredient, useful in the perspective to have 
ground states with non-abelian excitations, 
is given by the possibility of trapping, 
both for ultracold bosons and fermions, two 
(or more) hyperfine levels of the same atomic species. 

By combining the two ingredients, one may think to realize a system 
with two hyperfine levels, coupled between them and each one feeling 
a different artificial effective magnetic field. Several possible  
schemes have already been proposed in literature to realize such 
{\em non-abelian} gauge potentials, acting on the two (or more) components 
of an ultracold atomic gas: in \cite{ruse05} laser fields couple three internal 
states with a fourth (the so-called tripod scheme), in such a way 
that for two degenerate dark states one can have a truly 
non-abelian gauge potential. The extension of this scheme to the tetrapod 
setup has been recently discussed \cite{juze10}. Another proposed scheme 
is based on laser assisted tunneling depending on the hyperfine levels 
to obtain $U(2)$ vector potentials acting on ultracold atoms 
in optical lattices \cite{lewen05}, while effective non-abelian gauge 
potentials in cavity QED models has been recently addressed in 
\cite{larson09}. Several properties of ultracold atoms in artificial 
non-abelian gauge potentials, including 
Landau levels and dynamical regimes, have been recently studied  
\cite{clark06,santos07,lu07,clark08,larson09a,goldman09,lewen09,bermudez10}.

Motivated by these proposals, the natural question is 
whether one can find non-abelian anyons in such 
systems. Generally a non-abelian gauge potential in presence of strong interactions 
does not guarantee the non-abelianity of the excitations; 
rather, deformed Laughlin states (characterized by abelian quasi-holes) appear. 
Recently, different kinds of anomalous quantum Hall effect 
for ultracold atomic gases in artificial gauge potential 
have been addressed \cite{goldman09,lewen09,bermudez10}: in particular, 
in \cite{lewen09} it has been shown that a ultracold 
Fermi gas in a lattice subjected to a non-abelian gauge field can 
present non-chiral and anomalous quantum Hall effects.

In this paper we consider a non-abelian potential 
in a symmetric gauge acting on a two-component ultracold gas: 
this realistic potential (realizable with a suitable choice 
of the Rabi frequencies in a tripod scheme) 
has the advantage that Landau levels with high degeneracy are present 
and can be exactly determined. The two Landau 
levels become non-degenerate for a non-vanishing strength of the 
non-abelian potential: we show that, 
as expected, deformed Laughlin states arise. However, 
at the points in which there is degeneracy of the Landau levels, ground states 
with non-abelian excitations do appear and explicit analytical 
results can be obtained. These ground states, including deformed 
Moore-Read states (presenting Ising anyons), 
are studied for both fermionic and bosonic gases. 

\textit{Single-particle Hamiltonian:} We first consider the two-dimensional 
non-interacting Hamiltonian for atoms in two different hyperfine levels 
(hereafter denoted by $\left|\uparrow\right\rangle$ and $
\left|\downarrow\right\rangle$) in a 
non-abelian gauge potential $\vec A$:
\begin{equation} \label{ham}
H=\left( p_x+A_x\right)^2+\left( p_y+A_y\right)^2 \equiv H_a +H_{na}. 
\end{equation}
In the Hamiltonian (\ref{ham}) we explicitly introduced the (abelian) 
term proportional to the identity, $H_a$, and the (non-abelian) 
off-diagonal term $H_{na}$  (the units are taken to have mass $1/2$).

We choose the following vector potential $\vec A$: 
\begin{equation} \label{eq1}
A_x=q\sigma_x-\frac{B}{2}y \; , \qquad A_y=q\sigma_y+\frac{B}{2}x,
\end{equation}
where the $\sigma$'s are the $2 \times 2$ Pauli matrices. 
As we show in the Appendix, it is possible to engineer 
this vector potential in a system of rotating atoms having three 
quasi-degenerate ground states coupled with an excited state 
(the so-called tripod scheme): the vector potential $\vec{A}$ is in general 
determined by the choice of the parameters of the Rabi frequencies $\Omega_i$ 
\cite{ruse05}. In the Appendix we show that the 
Rabi frequencies in a rotating trap can be arranged to give the vector potential (\ref{eq1}). 

In (\ref{eq1}) the parameter $q$ measures the strength 
of the non-abelian contribution and $B$ is an artificial, diagonal 
magnetic field orthogonal to the system. 
When $q=0$, the usual Landau levels \cite{cooper08} are retrieved  
and are doubly degenerate. The total effective magnetic field is
\begin{equation} \label{eq1a}
 \mathcal{B}=\nabla \times \vec A + 
i \vec A\times\vec A= B - 2q^2 \sigma_z. 
\end{equation} 
One can see that the system is characterized by a 
translationally invariant Wilson loop \cite{goldman09}.

To diagonalize the Hamiltonian (\ref{ham}) we introduce the 
complex variables $z=x-iy$ and $\bar z =x+iy$, and we 
rewrite $H_a$ and $H_{na}$ as
\begin{equation} \label{eqhz}
H_a= 2q^2+B+\frac{1}{4}d^\dag d\, , \qquad
H_{na}= q \begin{pmatrix}0&-id\\id^\dag&0\end{pmatrix}
\end{equation}
where we introduced the creation and annihilation operators 
$d^\dag=B\bar z-4\partial_z$ and $d =Bz+4\partial_{\bar z }$.

The non-abelian part of the Hamiltonian $H_{na}$ corresponds to a Jaynes-Cummings coupling between subsequent Landau levels with different internal 
degree of freedom. Its spectrum is similar to the one obtained in the relativistic case typical of the graphene systems 
and, in particular, $H_{na}$ is analogous to the Hamiltonian obtained in \cite{lewen09} starting from the Dirac equation in an anisotropic regime. Therefore we can notice that it leaves a single family of uncoupled states, corresponding to the lowest Landau level, $\psi_{0}\left|\downarrow\right\rangle$, and, otherwise, its eigenstates take the form $\psi_{n-1}\left|\uparrow\right\rangle \pm \psi_{n}\left|\downarrow\right\rangle$.

\begin{figure}[width=12cm]
\centering
\includegraphics[width=12cm]{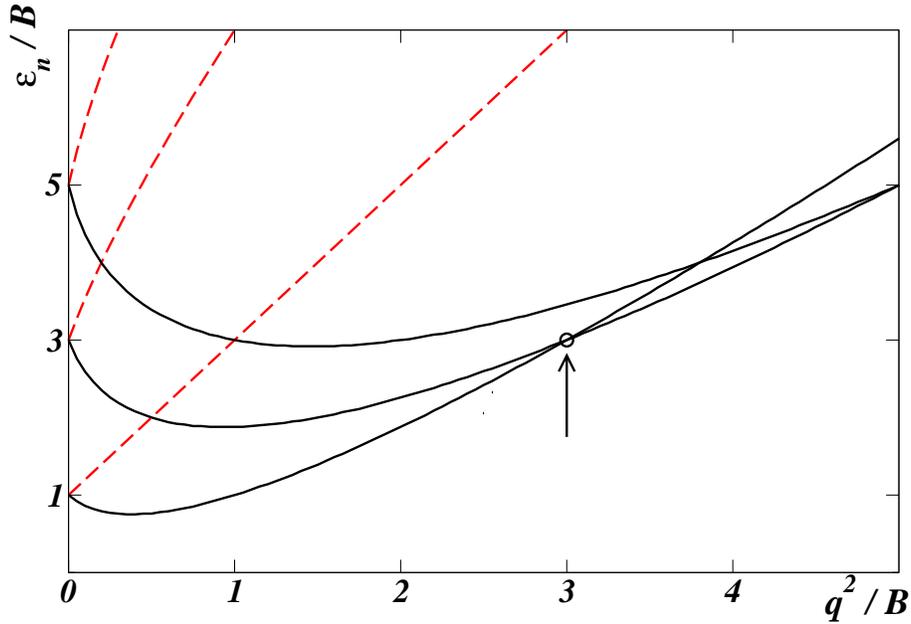}
\caption{Energies of the first $6$ eigenstates as 
a function of $q^2/B$. The $n^{\rm th}$ Landau level 
is split by the non-abelian contribution of the external field 
in $\chi_{n+1}^-$ (black solid) and $\chi_n^+$ (red dashed); 
for $q=0$ one recovers the usual Landau levels, while for $q^2/B=3$ 
there is a degeneracy of the Landau levels (indicated by a circle).}
\label{figlandau}
\end{figure}

Using these wavefunctions we can diagonalize the whole single-particle Hamiltonian $H$ obtaining the following eigenvalues
\begin{equation} \label{enlev}
 \varepsilon_n^{\pm}=2Bn+2q^2\pm\sqrt{B^2+8q^2Bn } 
\end{equation}
corresponding to the unnormalized eigenstates
\begin{equation} \label{eqeig}
 \chi_n^\pm=\left( B + 2q \sqrt{2Bn } \mp \sqrt{B^2+8q^2Bn}\right) \psi_{n-1}\left|\uparrow\right\rangle 
+ \left( B - 2q \sqrt{2Bn } \pm \sqrt{B^2+8q^2Bn}\right) \psi_{n}\left|\downarrow\right\rangle,
\end{equation}
where the $\psi_k$ are wavefunctions in the $k^{\rm th}$ Landau level such that 
$2\sqrt{2Bn } \psi_{n} = id^\dag \psi_{n-1}$, due to the Jaynes-Cummings coupling. 
Therefore there is a correspondence between the usual eigenstates in a 
magnetic field and the states $\chi$: 
every Landau level is split in two parts corresponding 
to the states $\chi_{n-1}^+$ 
and $\chi_{n}^-$ and for $q\to 0$ their energy is $\sim E_{n-1}\pm 4 q^2n$. 
The energy spectrum (\ref{enlev}) is plotted in Fig.\ref{figlandau}.

The uncoupled state family corresponds to $\chi_{0}^+$ and its energy, $B+2q^2$, is higher than the energy $\varepsilon_1^-$ of $\chi_1^-$, which is the ground state family of the system for $q^2<3B$ (the general case with $q^2\ge 3B$ will be analyzed in the following). To describe the states in the family $\chi_1^-$ it is useful to introduce the operator $\mathcal{G}_1 \equiv  c_{\uparrow,1} \sigma_x + c_{\downarrow,1} d^\dag$, where for $n=1,2\dots$ we defined the constants:
\begin{eqnarray*}
c_{\uparrow,n} &=& B+2q\sqrt{2Bn }+\sqrt{B^2+8q^2Bn } \\
c_{\downarrow,n} &=&i\left({B- 2q \sqrt{2Bn }-\sqrt{B^2+8q^2Bn }}\right)\left(2\sqrt{2Bn}\right)^{-1}.
\end{eqnarray*}
$\mathcal{G}_1$ allows to map uncoupled states in $\chi_0^+$ into states in $\chi_1^-$ so that we can describe every ground state in the form
\begin{equation}
\chi_1^-=\mathcal{G}_1 \left(P(z) e^{-\frac{B}{4}\left|z\right|^2}\left|\downarrow\right\rangle\right)
\end{equation}
where $P$ is a generic polynomial in $z$.

\textit{Two-body interactions:} 
We consider now a system of $N$ interacting atoms.
We will treat both fermions and bosons: 
in order to obtain quantum Hall states, a strong repulsion 
between atoms of the same hyperfine level is needed. 
For fermions the intra-species repulsion is provided by the Pauli principle; 
for bosons, we assume an (intra-species) 
interaction between $\left|\uparrow\right\rangle$ 
atoms or between $\left|\downarrow\right\rangle$ atoms of the form 
$H_I= g \sum_{i<j}^N \delta \left( z_i-z_j\right)$, where 
$g$ is a positive coupling constant such that $g\gg B,q$ 
(the $s$-wave scattering lengths between $\left|\uparrow\right\rangle$ 
atoms and between $\left|\downarrow\right\rangle$ atoms are assumed equal). 
For the sake of simplicity, we assume as well that the inter-species interactions 
(i.e., between $\left|\uparrow\right\rangle$ 
and $\left|\downarrow\right\rangle$ atoms) is vanishing: in the following 
we will also comment on the stability of the obtained ground states 
with respect to a non-vanishing repulsive inter-species interaction. 

In order to find a ground state of the multiparticle Hamiltonian $\mathcal{H}=\sum_i^N H_i + H_I$ we must exploit the single-particle ground state degeneracy to obtain a wavefunction where all the particles lie in a superposition of states in $\chi_1^-$ whose components $\left| \uparrow \uparrow \right\rangle_{ij}$ and $\left| \downarrow \downarrow \right\rangle_{ij}$ vanish if $z_i \to z_j$. For fermions one finds (for $q^2<3B$) that, given a Laughlin wavefunction
\begin{equation} \label{laughlin2}
\Lambda^{(m)}_N=  \prod_{i<j}^N \left(z_i - z_j \right)^m  e^{-\frac{B}{4}\sum_{i} ^N \left|z_i\right|^2} \left| \downarrow \downarrow ... \downarrow \right\rangle
\end{equation}
with $m$ odd, then the state 
\begin{equation} \label{eqgsb}
\Psi^{(m)} = \prod_j^N \mathcal{G}_{1;j}  \Lambda^{(m)}_N
\end{equation}
is a ground state of the Hamiltonian $\mathcal{H}$: 
every particle lies in a superposition of states $\chi_1^-$ and the antisymmetric wavefunction causes the intra-species interaction energy to be zero 
(here and in the following $\mathcal{G}_{n;j}$ denotes the 
operator $\mathcal{G}_n$ applied to the particle $j$). For repulsive bosons, 
the form (\ref{eqgsb}) holds with $m$ even and greater than 2 
(the states $\Psi^{(m)}$ with $m>2$ can be shown to be stable also
under the presence of an inter-species repulsive interaction).

The state $\Psi^{(m)}$ can be seen as a deformation, 
due to the non-abelian potential, of the usual Laughlin states and 
it describes an incompressible fluid of spin-$1/2$ particles:  
its norm can be easily written in terms of the norm 
of the corresponding Laughlin state,
showing that this state has, 
in the thermodynamical limit, a constant density. 
Also in the presence of quasi-holes in the Laughlin state
\begin{equation} 
 \Psi_{\zeta_1,\zeta_2}^{(m,k)}=\prod_i^N \mathcal{G}_{1;i} \left( z_i -\zeta_1\right)^k \left(z_i - \zeta_2 \right)^k \Lambda^{(m)}_N 
\end{equation}
the norm is related to the corresponding quantum Hall state: the Berry phase due to the adiabatic exchange of the quasi-holes $\zeta_1$ and $\zeta_2$ is the same of the one characterizing the corresponding quasi-holes in a simple Laughlin state. These excitations and their braiding statistics are therefore abelian \cite{arovas84}.

\begin{figure}[width=12cm]
\includegraphics[width=12cm]{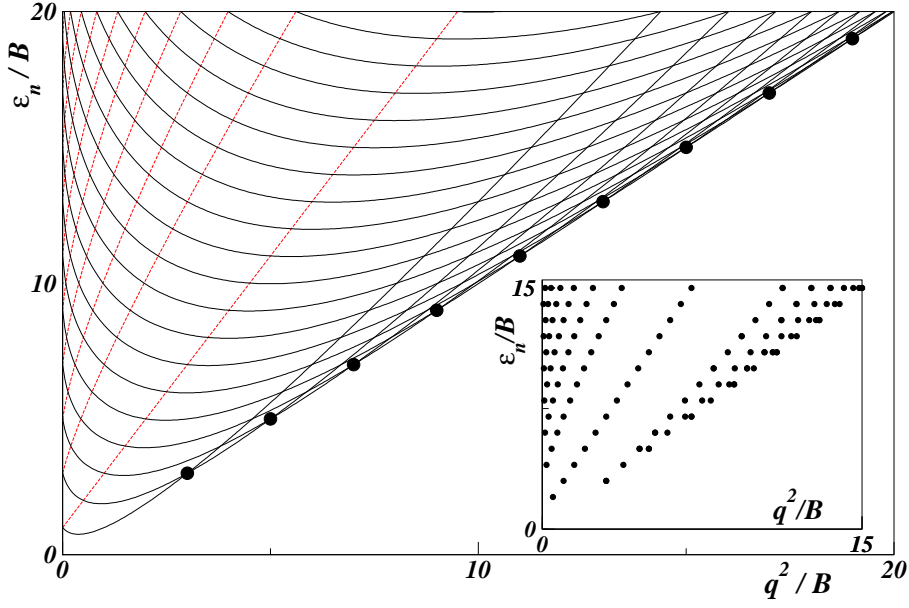}
\caption{Energies of $\chi_{n+1}^-$ (black solid line) and $\chi_n^+$ 
(red dashed line) for $n=0,\dots,10$ as a function of $q^2/B$. 
The crossings of the ground states happen in the degeneracy 
points $q^2/B=\left(1+2n \right)$, denoted by solid circles. Inset: 
Degeneracy points of Landau levels.}
\label{figgs}
\end{figure}

\textit{Higher values of $q^2/B$:} The previous results can be extended 
to $q^2/B>3$, where the ground state of the single-particle Hamiltonian 
is no longer $\chi_1^-$: one has that 
$\chi_{n>1}^-$ is the ground state for 
$\left(2n-1 \right)<q^2/B<\left(2n+1\right)$ (see Fig.\ref{figgs}). 
It is useful to introduce the operators 
$\mathcal{G}_n=c_{\uparrow,n}\,d^{\dag \, (n-1)}\sigma_x+c_{\downarrow,n}\,d^{\dag \,n}$
for every $n$ so that the ground state wavefunctions can be defined as 
$\chi_n^-=\mathcal{G}_n \left(P(z) e^{-\frac{B}{4}\left|z\right|^2}\left|\downarrow\right\rangle\right)$.

Using these expressions it is possible to obtain, following the procedure shown in the case of $\chi_1^-$, the appropriate many-body wavefunctions for every value of $B$ and $q$. In particular all the antisymmetric states given by
\begin{equation} \label{eqgsc}
 \Psi_n^{(m)} = \prod_j^N \mathcal{G}_{n;j}  \Lambda^{(m)}_N
\end{equation}
(with $\Lambda^{(m)}_N$ an odd Laughlin state (\ref{laughlin2})) are fermionic ground states. Besides, for even values $m>2n$, also the symmetric bosonic wavefunction $\Psi_n^{(m)}$ describes a possible ground state for both intra-species and inter-species interactions.

\textit{Degeneracy points and non-abelian anyons:} 
An important property of the spectrum (\ref{enlev}) is that there are 
points in which degeneracies of the Landau levels arise: 
for $q^2/B=1+2n$ the two lowest energy levels cross and 
the ground state degeneracy of the single particle is doubled. Notice 
that there is complete sequence of 
points in which two Landau levels have the same energy (integer in unit of $B$) as depicted in the inset of Fig. \ref{figgs}.

For $q^2/B=1+2n$, the single-particle ground states 
are all the superpositions of wavefunctions in $\chi_n^-$ and $\chi_{n+1}^-$. 
Let us consider for simplicity the case of $q^2=3B$ at the crossing between the states $\chi_1^-$ and $\chi_2^-$ 
(but our conclusions can be easily generalized to all the other 
degenerate points). Once we introduce 
the intra-species interaction between atoms, 
this double degeneracy implies a novel form of the multiparticle ground 
state which is quite different from (\ref{eqgsb}): 
in the degeneracy points the antisymmetrization hides a clustering 
of the particles into two sets of $N$ particles $A$ and $B$ that are physically different and refer to states $\chi_1^-$ and $\chi_2^-$. For fermions, we can rewrite the ground state $\Omega_c$ for $2N$ 
particles making explicit this clustering:
\begin{equation} \label{eqcluster}
 \Omega_c=\mathcal{A}\left[ \prod \limits_{k \in A} \mathcal{G}_{1;k} \prod \limits_{i<j \in A} \left( z_i - z_j\right)   \prod \limits_{l \in B} \mathcal{G}_{2;l} \prod \limits_{i<j \in B} \left( z_i - z_j\right)      \right] e^{-\frac{B}{4}\sum \limits_{i} ^{2N} \left|z_i\right|^2} \left| \downarrow \downarrow ... \downarrow \right\rangle.  
\end{equation}
$\Omega_c$ is obtained through a Slater determinant involving all the states $\chi_1^-$ and $\chi_2^-$ up to the power $z^{N/2 -1}$. To create a quasi-hole excitation, the total angular momentum must be increased by $N/2$ and, given $2M$ quasi-holes of coordinates $\zeta_h$, the corresponding wavefunction can be expressed as 
\begin{multline} \label{eqholes2} 
 \Omega_{qh}=\mathcal{A}\left[ \prod\limits_{A} \mathcal{G}_{1;k}  \prod\limits_{B} \mathcal{G}_{2;l} \prod \limits_{A}  \left( z_i - z_j\right)  \prod\limits_{A}\prod\limits_{h=1}^{M}\left(z_i-\zeta_h \right) \right. \\ \left. \prod \limits_{B} \left( z_i - z_j\right) \prod\limits_{B}\prod\limits_{h=M+1}^{2M}\! \! \! \left(z_i-\zeta_h \right)    \right] e^{-\frac{B}{4}\sum \limits_{i} ^{2N} \left|z_i\right|^2} \left| \downarrow ... \downarrow \right\rangle
\end{multline}
These quasi-holes obey a fermionic statistics once two of them in the same cluster ($A$ or $B$) are exchanged; however they show
the same fusion rules of the Ising model with defined fermionic parity \cite{georgiev09} characterizing the Moore and Read (MR) Pfaffian state \cite{moore91,nayak96}. We notice that if one is allowed to consider their linear superpositions at the degeneracy
points of this ultracold atoms setup (corresponding to atoms in the space spanned by $\chi_1^-$ and $\chi_2^-$ 
moving in higher angular momentum states), then such superpositions
obey non-Abelian braiding rules.

The state $\Omega_c$ is the ground state corresponding to the highest possible density in the degeneracy point given by $q^2=3B$; decreasing the filling factor (i.e. the density) the ground states are characterized by polynomials with higher degrees; thus it is possible to obtain other ground states of the system with $2N$ particles different from (\ref{eqgsb}) and (\ref{eqcluster}), but nevertheless characterized by a clustering of the particles in $\chi_1^-$ and $\chi_2^-$. We can introduce, for instance, the wavefunctions:
\begin{equation} \label{omega} 
 \Omega^m_{\rm Hf} = {\rm Hf}\left( \left( {\mathcal G}_{1;i}
{\mathcal G}_{2;j}-{\mathcal G}_{2;i} {\mathcal G}_{1;j}\right) \frac{1}{z_i-z_j}\right) \Lambda^{(m)}_{2N} 
\end{equation}
where $m$ is an odd integer, $\Lambda^{(m)}_{2N}$ is the Laughlin state (\ref{laughlin2}) for $2N$ particles, 
and $\rm Hf$ indicates the Haffnian, which is a symmetric version of the Pfaffian. The state $\Omega^m_{\rm Hf}$ is antisymmetric over all the particles and this guarantees that the inter-species interaction gives a zero contribution, even if the wavefunction $\Omega^1_{\rm Hf}$ doesn't vanish for $z_i \to z_j$, that makes it closely related to the MR state. Moreover, for every $m>4$ (also for even $m$), $\Omega^m_{\rm Hf}$ is a ground state for every two-body repulsive contact interaction, therefore it is stable under (repulsive) inter-species perturbations.

Another interesting ground state of $\mathcal{H}$ (with higher angular momentum) is a deformed 
MR state that can be described by the wavefunction:
\begin{equation} \label{eqmr}
 \Omega_{\rm MR}= \mathcal{S}\left( \prod\limits_{i=1}^N \mathcal{G}_{1;i} \prod\limits_{i=1+N}^{2N} \mathcal{G}_{2;i} \right){\rm Pf}\left( \frac{1}{z_i-z_j}\right)  \Lambda^{(2)}_{2N}
\end{equation}
where $\mathcal{S}$ is the symmetrization over all the particles and $\rm Pf$ is the Pfaffian operator. This state shares all the main characteristics of the MR wavefunction \cite{nayak96}, and, in particular, its excitations, introduced through a clustering similar to equation (\ref{eqholes2}), are non-abelian Ising anyons. Notice that the state (\ref{eqmr}) is formally identical to a state 
considered in \cite{nayak96}, where it was shown that it has the same 
statistical properties of the MR state. 
In fact, the relation between $\Omega_{\rm MR}$ and the usual MR state \cite{moore91,nayak96} allows to relate their norms; therefore, the exchange statistics characterizing excitations $\zeta$ of the kind
\begin{multline*} 
 \Omega_{\rm MR}\left(\zeta_a,\zeta_b, \zeta_c, \zeta_d \right) = \mathcal{S}\left( {\prod}_{i=1}^N \mathcal{G}_{1;i} {\prod}_{i=1+N}^{2N} \mathcal{G}_{2;i} 
\right) \\
{\rm Pf}\left( \frac{\left(z_i-\zeta_a \right) \left(z_i-\zeta_b \right)\left(z_j-\zeta_c \right) \left(z_j-\zeta_d \right) + i\leftrightarrow j }{z_i-z_j}\right) \Lambda^{(2)}_{2N}
\end{multline*}
is described by four Ising anyons as in the case of the MR state, because the monodromy and Berry phases acquired in the exchange of two excitations coincide.

\textit{Conclusions:} 
In this paper we studied ultracold atoms in an artificial non-abelian gauge 
potential: we computed the single-particle energy levels, showing 
that the non-abelian part of the vector potential splits the degeneracy 
of Landau levels. In presence of strong intra-species interactions, deformed 
Laughlin states and abelian excitations are obtained. However, at the 
points in which the Landau levels are again degenerate, 
multiparticle ground states characterized by a clustering 
similar to the one identified in Moore and Read states emerge. 
The system is then suitable to have non-abelian anyonic excitations, 
as the Ising anyons, opening the possibility to tune 
artificial non-abelian gauge potentials in ultracold systems 
to induce and manipulate non-abelian anyons.  

Given the importance of non-abelian anyons for topological quantum computation, 
an important question for future work is the study of the braiding rules 
for the non-abelian anyons obtained at the degeneracy points: 
from this point of view, one could consider more components (e.g., with tetrapod schemes) 
in order to explore the possibility to have lines or eventually regions 
of degeneracy. 
Similarly interesting would be the characterization of the topological order 
\cite{wen95} in the transitions between deformed Laughlin states and deformed 
Moore-Read states at the degeneracy points, which could show similarities 
with transitions between Halperin and Moore-Read states in bilayer systems \cite{peterson10}.

\textit{Acknoledgements.} 
We thank A. Cappelli, A. Marzuoli, G. Mussardo, K. Schoutens, 
X. Wan and D. Fioretto for very useful discussions. This work is supported 
by the grants INSTANS (from ESF) and 2007JHLPEZ (from MIUR).

\appendix

\section{Artificial non-abelian gauge potential in a rotating tripod system} \label{app}

The aim of this Appendix is to show how it is possible to obtain the following $2\times 2$ non-abelian potential
\begin{equation} \label{eq1-app}
A_x=q\sigma_x-\frac{B}{2}y\mathbb{I} \; , \qquad A_y=q\sigma_y+\frac{B}{2}x\mathbb{I}
\end{equation}
starting from a fine tuned system of tripod atoms of 
the kind described in \cite{ruse05} (here $\sigma_x$ and $\sigma_y$ are the usual Pauli matrices and $\mathbb{I}$ denotes the identity matrix). In particular we will obtain the potential $A(q)$ starting from a particular family of configurations of the Rabi frequencies $\Omega_\mu$ in \cite{ruse05}:
\begin{equation}
 \Omega_1 = \varOmega \sin\left( \theta\right)\cos\left(\phi\right)e^{iS_1}\,,\quad \Omega_2 = \varOmega \sin\left( \theta\right)\sin\left(\phi\right)e^{iS_2}\,,\quad \Omega_3=\varOmega\cos\left(\theta\right)e^{iS_3}  
\end{equation}
where $S_1$ and $S_2$ are functions of the position and of the parameter $q$ while the angles $\phi$ and $\theta$ and $S_3$ are chosen constants.

The key elements to obtain the desired vector potential 
$\vec A$ (\ref{eq1-app}) are the rotation of the whole system (in a way similar to the rotating system in \cite{lu07}) and a suitable gauge transformation. 
Let us consider a system of tripod atoms in an inertial frame of reference characterized by a non-abelian gauge potential $\tilde A$, a scalar potential $V_{\rm rot} \equiv \Phi(\tilde A) + V$ as the one described in 
\cite{ruse05} and a harmonic confining potential $\omega r^2 / 4$ (here and in the following we will consider $m=1/2$ and $\hbar=1$). The corresponding Hamiltonian reads:
\begin{equation}
 H_{\rm IF}=\left( p+\tilde A\right) ^2 + \frac{1}{4}\omega^2 r^2  + V_{\rm rot}
\end{equation}
If we put the whole system in rotation with angular velocity $\Omega$ the Hamiltonian in the rotating frame of reference reads:
\begin{equation}
 H_{\rm Rot}=\left( p+\tilde A\right) ^2 + \frac{1}{4}\omega^2 r^2 +\Omega \mathbb{L}_z  + V_{\rm rot}
\end{equation}
where we introduced the gauge invariant angular momentum
\begin{equation}
\mathbb{L}= \vec r \times \left(\vec p + \tilde A \right)
\end{equation}
and all the coordinates are now considered in the rotating frame.
We can now rewrite $H_{\rm Rot}$ introducing the gauge potential:
\begin{equation} \label{eq2}
A_x= \tilde A_x -\frac{B}{2}y\,,\qquad A_y = \tilde A_y +\frac{B}{2}x 
\end{equation}
where we put $B=\omega$. We obtain:
\begin{equation} \label{eqH2}
 H_{\rm Rot}=\left( p+A\right) ^2 +  \Delta  \mathbb{L}_z  + V_{\rm rot} 
\end{equation}
with $\Delta = \Omega - \omega$.

Our aim is to identify the correct family of Rabi frequencies, $V_{\rm rot}$ and gauge transformation such that:
\begin{equation} \label{eqH}
 H=\left( p+A\right) ^2 +  \Delta  L_z  
\end{equation}
with $A$ given by (\ref{eq1-app}) and $L_z=\vec r \times \vec p$ being the usual angular momentum in the rotating frame. In particular we need:
\begin{eqnarray}
 \tilde A &=& \left(q\sigma_x,q\sigma_y\right) \\
V_{\rm rot} &=& q \Delta (y\sigma_x - x\sigma_y)
\end{eqnarray}


In order to obtain, in the rotating frame, the potential $A$ in (\ref{eq1-app}) starting from the potential $\mathcal{A}$ given in \cite{ruse05}
\begin{eqnarray}
 \mathcal{A}_{11} &=& \cos^2 \phi \nabla S_{23}+\sin^2\phi \nabla S_{13} \\
 \mathcal{A}_{22} &=& \cos^2 \theta \left( \cos^2\phi \nabla S_{13}+\sin^2\phi \nabla S_{23}\right)  \\
 \mathcal{A}_{12} &=& \cos \theta \left( \frac{1}{2} \sin 2\phi \nabla S_{12}-i\nabla\phi\right) 
\end{eqnarray}
we need a suitable unitary gauge transformation $O(\vec r)$. In particular the field transforms as $\mathcal{A} \to O  \mathcal{A} O^\dag -i O \nabla O^\dag$ and thus we must have
\begin{equation}
 O \mathcal{A} O^\dag -i O \nabla O^\dag = \tilde A = \left(q \sigma_x, q \sigma_y\right) 
\end{equation}

From the definition of $\mathcal{A}$ it is evident that, choosing a constant $\phi$, it is not possible to obtain $\mathcal{A}_y \propto \sigma_y$ but it is easy to check that we can obtain $\mathcal{A}_y = k \mathbb{I} - q\sigma_z$ and $\mathcal{A}_x = q\sigma_x$ for a suitable choice of the parameters as functions of $q$. Therefore the gauge transformation we will apply is:
\begin{equation} \label{gauge}
 \Psi \to O \Psi \,,\qquad {\rm with} \; O=e^{iky-i\frac{\pi}{4}\sigma_x}
\end{equation}
with $k$ to be defined in the following. In this way we obtain:
\begin{equation} \label{eqa1}
 \mathcal{A}_x = q \sigma_x \xrightarrow{O} \tilde A_x = q\sigma_x \,,\qquad \mathcal{A}_y = k \mathbb{I} - q\sigma_z \xrightarrow{O} \tilde A_y = q\sigma_y
\end{equation}
We must also consider that the scalar potential in \cite{ruse05} 
are affected by $O$ as $O\left( V+\Phi \right)O^\dag$, thus, in order to obtain (\ref{eqH}) out of (\ref{eqH2}) we must have:
\begin{equation} 
 O\left( V+\Phi \right)O^\dag = V_{\rm rot} = -\Delta \vec r \times \tilde A
\end{equation}
and then:
\begin{equation} \label{eqa2}
 V+\Phi = q\Delta\left( y \sigma_x + x \sigma_z\right) 
\end{equation}


Now we can find the suitable parameters to satisfy (\ref{eqa1}) and (\ref{eqa2}). First of all we impose $\phi=\pi/4$ and $S_3={\rm cost}$. Then, from the definition of $\mathcal{A}$ we obtain that:
\begin{eqnarray}
 \partial_x \left( S_{1} + S_{2}\right)  =  0 \\
 \cos \theta \partial_x \left( S_1 - S_2 \right) =2q \\
2k - 2q = \partial_y \left( S_{1} + S_{2}\right) \\
2q + 2k =\cos^2 \theta \partial_y \left( S_{1} + S_{2}\right)
\end{eqnarray}
A possible solution is given by:
\begin{eqnarray}
 S_1 &=& \lambda\left( x + y\right)  \\
 S_2 &=& \lambda\left( -x + y\right) 
\end{eqnarray}
with $\lambda = q/\cos\theta$ in order to satisfy the first two equations; we obtain from the last two equations:
\begin{equation}
 \cos^2 \theta-2\cos \theta -1 = 0 \quad \Rightarrow  \quad \cos \theta = 1 - \sqrt{2}
\end{equation}
Therefore:
\begin{equation} \label{eqk}
 \lambda = \frac{q}{\cos\theta}=\frac{q}{1-\sqrt{2}} \, , \qquad k = \frac{1+\cos^2\theta}{2\cos\theta}\,q=\frac {2 - \sqrt{2}}{1 - \sqrt{2}} \,q
\end{equation}

In this way we found the values of $S_1,S_2,\phi,\theta$ in order to obtain the right $\tilde A$ after the gauge transformation.
Let us consider now the scalar potentials; imposing $\phi = \pi/4$ and $\cos\theta=1-\sqrt{2}$ we find from \cite{ruse05} and from (\ref{eqa2}):
\begin{eqnarray}
 V_{11}+\Phi_{11}&=&\frac{V_1+V_2}{2}+\frac{\lambda^2}{4}\sin^2\theta=  q \Delta x \\
 V_{22}+\Phi_{22}&=&\frac{V_1+V_2}{2}\cos^2\theta+V_3\sin^2\theta + \lambda^2\cos^2\theta\sin^2\theta = -q \Delta x\\
 V_{12}+\Phi_{12}&=&\frac{V_1-V_2}{2}\cos\theta=q \Delta y
\end{eqnarray}
The solution is given by:
\begin{eqnarray}
 V_1&=&q \Delta x + \lambda \Delta y - \frac{\lambda^2}{4} \sin^2\theta  \label{v1}\\
 V_2&=&q \Delta x - \lambda \Delta y - \frac{\lambda^2}{4} \sin^2\theta \label{v2}\\
 V_3&=&k \Delta x -\frac{3}{4}q^2 \label{v3}
\end{eqnarray}
with $\lambda$ and $k$ given by (\ref{eqk}).


In conclusion, we showed that the values of the angles
\begin{equation}
 \phi=\frac{\pi}{4} \qquad \theta=\arccos\left(1-\sqrt{2} \right) 
\end{equation}
and the linear functions $S_1,S_2,V_1,V_2,V_3$ described above, allow to define the right Rabi frequencies in order to obtain, through the gauge transformation (\ref{gauge}), the Hamiltonian (\ref{eqH}) for a rotating system. In particular we are able to describe a family of rotating physical systems, characterized by $q$, with the potential $A(q)$ (\ref{eq1-app}) that has been analyzed in our paper.

\end{document}